# First-principles calculations of the structural, electronic, optical and elastic properties of the CuYS$_2$ semiconductor

M.G. Brik[*]

Institute of Physics, University of Tartu, Riia 142, Tartu 51014, Estonia

**Abstract**

Ternary semiconductor CuYS$_2$ is studied by using the first-principles methods in the density functional theory (DFT) framework. The structural, electronic, optical and elastic properties were calculated at the ambient and elevated hydrostatic pressures. The compound was shown to have an indirect band gap of about 1.342/1.389 eV (in the generalized gradient and local density approximations). The anisotropy of the optical properties was studied by calculating the absorption spectra, dielectric function and index of refraction for different polarizations. The anisotropy of the elastic properties was visualized by plotting the three-dimensional dependence of the Young's moduli on the direction in the crystal lattice. The obtained results, which are reported for the first time to the best of the author's knowledge, can facilitate assessment of possible applications of the title material.

## 1. Introduction

For a long time, the I-III-VI$_2$ ternary semiconductors (I=Cu, Ag; III=Al, Ga, In, Sc, Y, La; VI=S, Se, Te) have been a subject of thorough experimental and theoretical investigations because of their applications in non-linear optics and solar cell industry [1, 2, 3, 4, 5 etc]. The main emphasis so far was put on the so called Group IIIA compounds, with III=Al, Ga, In, which crystallize in the chalcopyrite structure, whereas the IIIB materials group (III=Sc, Y, La), whose crystal lattices have, as a rule, a lower symmetry, remain much less explored. Thus, the structure of $CuScS_2$ and its electronic properties were reported in Refs. [6,7]. The structural properties of $CuYS_2$ were described in Refs. [8, 9], but so far, to the best of the author's knowledge, no theoretical and experimental studies of the electronic, optical, elastic properties for this compound can be found in the literature. Therefore, in the present paper the results of such calculations, performed for the ambient and elevated hydrostatic pressure in the pressure range from 0 to 20 GPa, are reported and discussed. The obtained information can be useful for assessing perspectives of potential applications of $CuYS_2$; it can be compared in the future with the experimental results, should the corresponding measurements be performed.

---

[*] Corresponding author. E-mail: brik@fi.tartu.ee

The paper is organized as follows: in the next section the structure of $CuYS_2$ and method of calculation are described. After that, the paper is continued with presenting the results of the structural, electronic, optical and elastic properties calculations, before it is concluded with a short summary.

## 2. Crystal structure of $CuYS_2$ and method of calculations

According to Ref. [9], $CuYS_2$ crystallizes in the Pnma space group (No. 62), with four formula units in one unit cell. The lattice parameters are collected in Table 1. The Cu ions are four-fold coordinated by the S ions, whereas the Y ions are six-fold coordinated by the S ions. There are two kinds of sulfur ions in this structure. The first one is in the octahedral coordination, with three Y and three Cu ions as the nearest neighbors, and the second one is in the tetrahedral coordination having three Y and one S ion in the first coordination sphere.

The structural data from Ref. [9] were taken as an initial input for all calculations of the structural, electronic, optical and elastic properties of $CuYS_2$. The CASTEP module [10] of Materials Studio was used with either generalized gradient approximation (GGA) with the Perdew–Burke–Ernzerhof functional [11] or the Ceperley–Alder–Perdew–Zunger parameterization [12, 13] in the local density approximation (LDA) to treat the exchange–correlation effects.

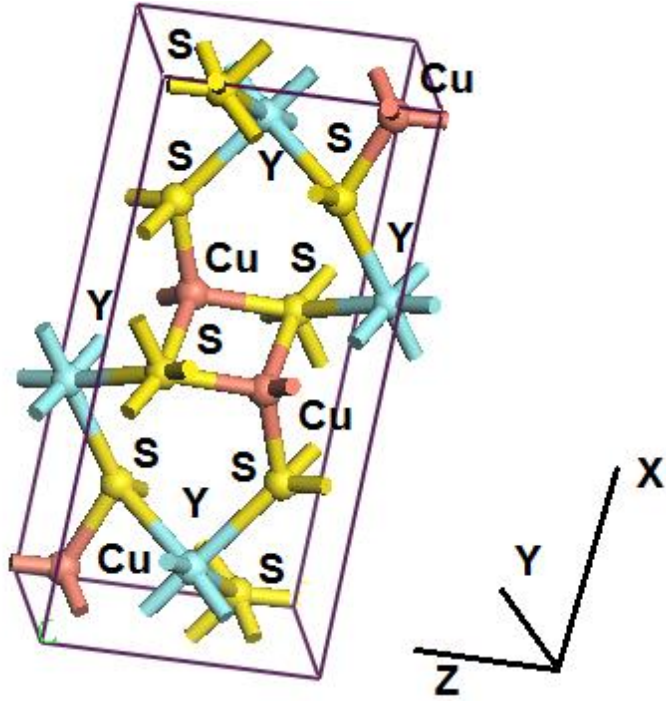

Fig. 1. One unit cell of $CuYS_2$.

The plane-wave basis set cut-off was set at 290 eV; the Monkhorst–Pack k-points grid sampling was set at 2×8×4 points for the Brillouin zone. Such a choice of the k-point set corresponded to the following separation between them in the inverse space: 0.036 Å$^{-1}$, 0.031 Å$^{-1}$, and 0.039 Å$^{-1}$ along the reciprocal lattice coordination axes $q_1$, $q_2$, $q_3$. The convergence tolerance parameters were: energy $10^{-5}$ eV, force 0.03 eV/Å; stress 0.05 GPa; displacement 0.001 Å. The electronic configurations were $3d^{10}4s^1$ for Cu, $4s^24p^64d^15s^2$ for Y, $3s^23p^4$ for S.

## 3. Results of calculations

3.1. Structural properties

The crystal structure data for $CuYS_2$, optimized at the ambient pressure with the above-given calculating settings in comparison with the experimental data from Ref. [9], are collected in Table 1.

Table 1. Experimental and calculated crystal structure parameters, including fractional coordinates of all ions in a unit cell, for $CuYS_2$ at ambient pressure

|  | Experim. [9] | Calculated (this work) | |
| --- | --- | --- | --- |
|  |  | GGA | LDA |
| $a$, Å | 13.453 | 13.7584 | 13.1964 |
| $b$, Å | 3.9812 | 4.0127 | 3.8582 |
| $c$, Å | 6.2908 | 6.3517 | 6.1114 |
|  | 0.4465 | 0.45454 | 0.4587 |
| Cu | 0.250 | 0.250 | 0.250 |
|  | 0.3919 | 0.37208 | 0.36863 |
|  | 0.13488 | 0.13213 | 0.13233 |
| Y | 0.250 | 0.250 | 0.250 |
|  | 0.4925 | 0.49665 | 0.5011 |
|  | 0.4623 | 0.46118 | 0.45891 |
| S1 (octahedral) | 0.250 | 0.250 | 0.250 |
|  | 0.7597 | 0.74572 | 0.74805 |
|  | 0.2934 | 0.29819 | 0.29761 |
| S2 (tetrahedral) | 0.250 | 0.250 | 0.250 |
|  | 0.2310 | 0.24614 | 0.24894 |

As seen from this Table, agreement between the calculated and experimental crystal lattice parameters is good. In addition, the theoretical and experimental fractional coordinates of all ions in a unit cell also match each other well, which serves as an additional proof of reliability of these theoretical findings and gives confidence in the results of the following calculations of the electronic, optical and elastic properties of $CuYS_2$ presented in the next sections.

3.2. Electronic properties

The calculated band structure of $CuYS_2$ is shown in Fig. 2, whereas Fig. 3 shows the Brillouin zone of $CuYS_2$ with indication of a path along which the cross-section of the energy surfaces is shown in Fig. 2.

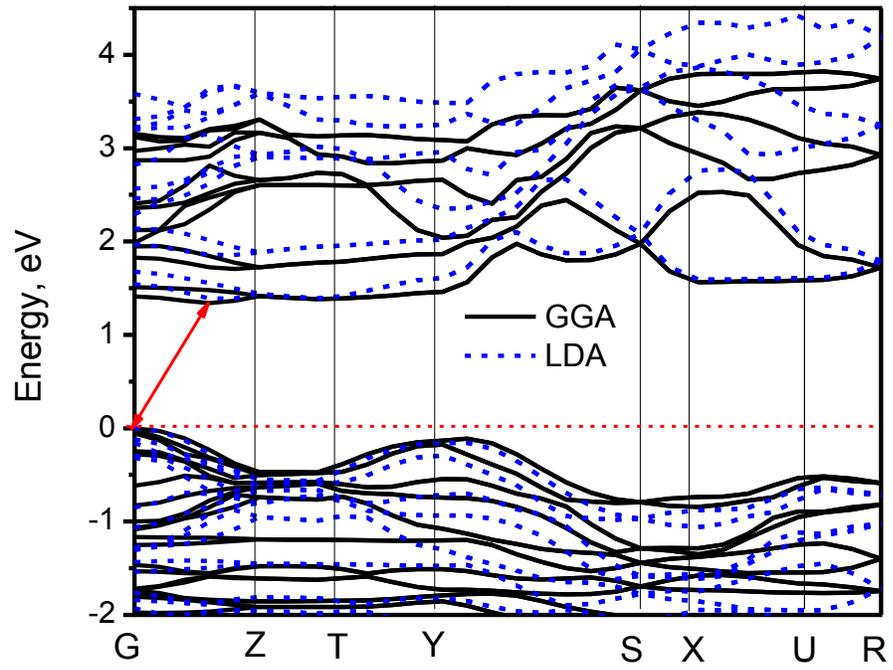

Fig. 2. Calculated band structure of CuYS$_2$. Indirect band gap is shown by an arrow.

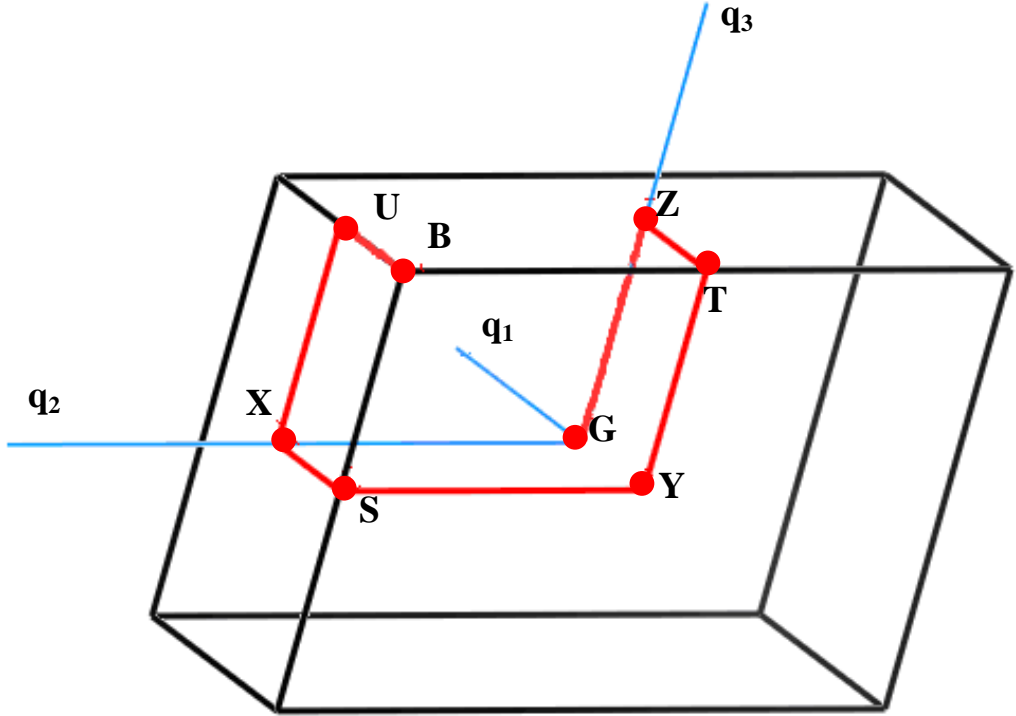

Fig. 3. Brillouin zone of CuYS$_2$. The red line corresponds to the path of the band structure diagram.

The calculated band gap is 1.342 eV (GGA) and 1.389 eV (LDA); it turns out to be of an indirect character, since the maximum of the valence band (VB) is realized at the G point (the Brillouin zone center), whereas the minimum of the conduction band (CB) is in the segment between the G and Z points. It is well known that both GGA and LDA underestimate the calculated band gaps; therefore, the above-given numbers can serve as the lower estimates of the true band gap, whose actual experimental value can be about 1 – 1.5 eV greater, i.e. between 2.3 and 2.5 eV. Unfortunately, no data on the experimental measurements of the band gap for the title compound were reported so far. An argument in favor of this estimate is the experimental band gap value of a similar CuScS$_2$ compound, which was given as 2.3 eV [8], and the theoretical indirect band gap of CuScS$_2$ 1.99 eV [7]. Both CB and VB states are not localized in energy and exhibit well-pronounced dispersion.

The composition of the calculated band gaps can be deduced from the density of states (DOS) diagrams, presented in Fig. 4. The CB, whose width is about 3 eV, consists mainly of the Y 4d states, with a contribution of the 4s states of Cu. The dominating states in the VB (its width is about 6 eV) come from completely filled sulfur 3p states and Cu 3d states. These S 3p and Cu 3d states are strongly mixed with each other (highly hybridized). Quite noticeable difference between the distributions of the 3p states of the octa- and tetrahedrally coordinated S ions can be observed. The 3s states of the tetrahedral sulfur ions are somewhat higher (at -12 eV) than the 3s states of the octahedral sulfur ions (at -13 eV). Finally, deep 4p and 4s states of Y produce narrow bands sharply peaked at -22 eV and -44 eV, respectively. Strong hybridization between the S and Cu states leads to high covalency of the chemical bonds between them. This conclusion can be confirmed by comparing the calculated effective Mulliken charges of all ions. They are as follows: (in units of the proton charge, the GGA/LDA values are given): Cu 0.06/-0.1; Y 0.90/0.91; S(tetr.) -0.49/-0.43; S(oct.) -0.48/-0.38. Very big difference from the formal charges expected from the chemical formula can be easily noticed.

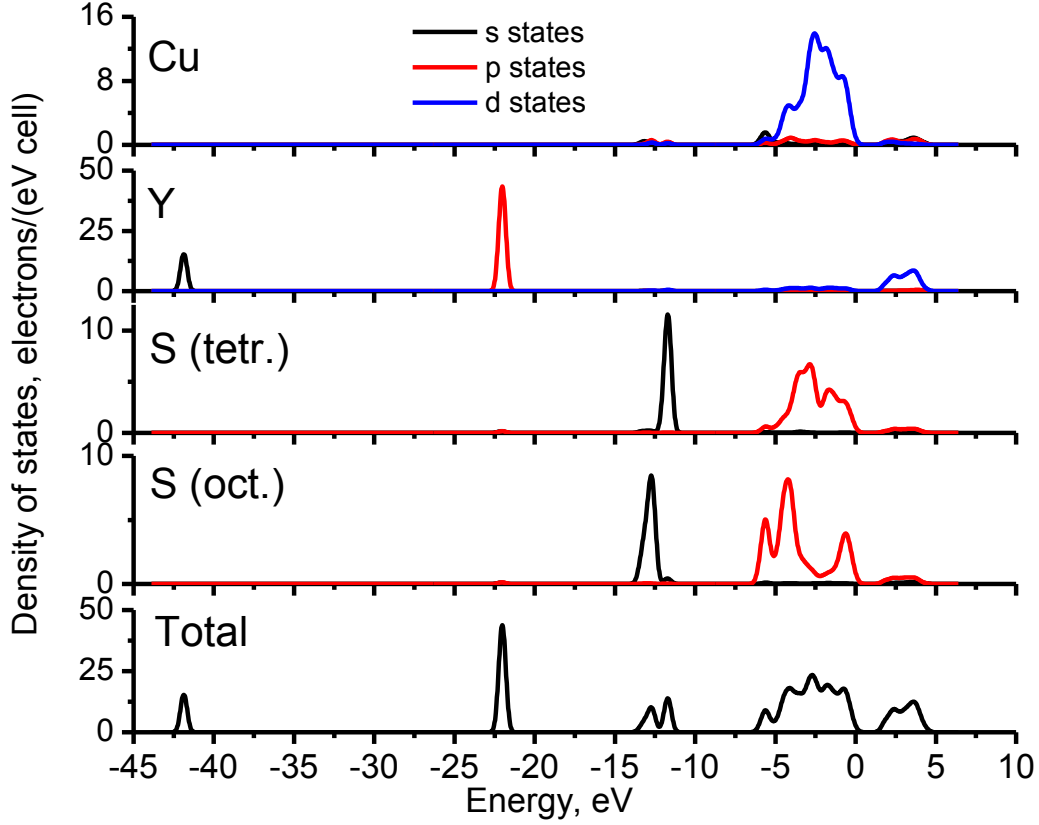

Fig. 4. Calculated DOS diagrams for CuYS$_2$.

3.3. Optical properties

The optical properties of a solid can be effectively described in a unique way by its complex dielectric function $\varepsilon$. After calculations of the electronic structure, the optical properties can be calculated in CASTEP in a straightforward manner. The imaginary part Im($\varepsilon(\omega)$) of a dielectric function $\varepsilon(\omega)$ is calculated by direct numerical evaluations of the matrix elements of the electric dipole operator between the occupied states in the VB and empty states in the CB:

$$\mathrm{Im}\left(\varepsilon(\omega)\right) = \frac{2e^2\pi}{\omega\varepsilon_0} \sum_{\mathbf{k},v,c} \left|\left\langle \Psi_{\mathbf{k}}^c \left| \mathbf{u}\cdot\mathbf{r} \right| \Psi_{\mathbf{k}}^v \right\rangle\right|^2 \delta\left(E_{\mathbf{k}}^c - E_{\mathbf{k}}^v - E\right), \tag{1}$$

where **u** is the polarization vector of the incident electric field, **r** and $e$ are the electron's radius-vector and electric charge, respectively, $\Psi_{\mathbf{k}}^c$, $\Psi_{\mathbf{k}}^v$ are the wave functions of the CB and VB, respectively, $E = \hbar\omega$ is the incident photon's energy, and $\varepsilon_0$ is the dielectric permittivity of vacuum. The summation in Eq. (1) is carried out over all states from the occupied and empty bands, with their wave functions obtained in a numerical form after optimization of the crystal structure. The imaginary part Im($\varepsilon(\omega)$) is proportional to the absorption spectrum of a solid. It should be kept in mind that Eq. (1) invokes certain approximations. For example, the local field effects (influence of the polarizability of a crystal onto the electric field inside the material) are not taken into account. In addition, the excitonic effects are also not considered. Moreover, an intrinsic error in the matrix elements for optical transitions exists due to the fact that the pseudo-wave functions are used in the calculations, which deviate from the true wave functions in the core region (Ref. [14] and references therein).

The real part Re($\varepsilon(\omega)$) of the dielectric function $\varepsilon$, which determines the dispersion properties and refractive index values, is estimated then by using the Kramers-Kronig relation:

$$\mathrm{Re}(\varepsilon(\omega)) = 1 + \frac{2}{\pi} \int_0^\infty \frac{\mathrm{Im}(\varepsilon(\omega'))\omega' d\omega'}{\omega'^2 - \omega^2} \qquad (2)$$

Figs. 5 and 6 show the GGA - and LDA -calculated dielectric function $\varepsilon$ (both real and imaginary parts).

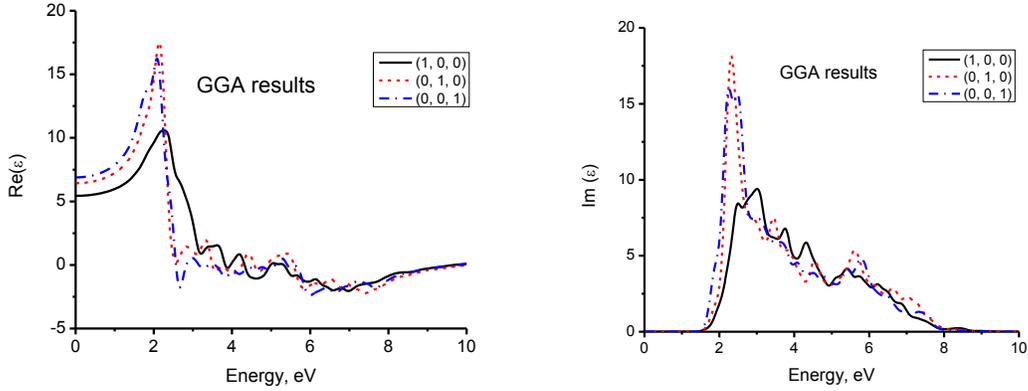

Fig. 5. The GGA-calculated real Re($\varepsilon$) and imaginary Im($\varepsilon$) parts of dielectric function $\varepsilon$ for CuYS$_2$.

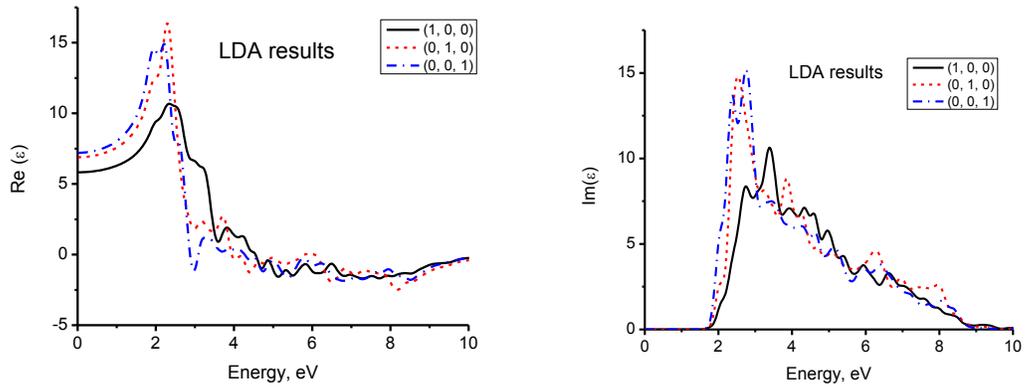

Fig. 6. The LDA-calculated real Re($\varepsilon$) and imaginary Im($\varepsilon$) parts of dielectric function $\varepsilon$ for CuYS$_2$.

The square root of Re($\varepsilon$) in the limit of infinite wavelengths gives an estimation of the refractive index *n* of a solid. Such estimates are as follows (the GGA/LDA values are given): $n$(1, 0, 0)=2.33/2.41; $n$(0, 1, 0)=2.54/2.62; $n$(0, 0, 1)=2.62/2.68.

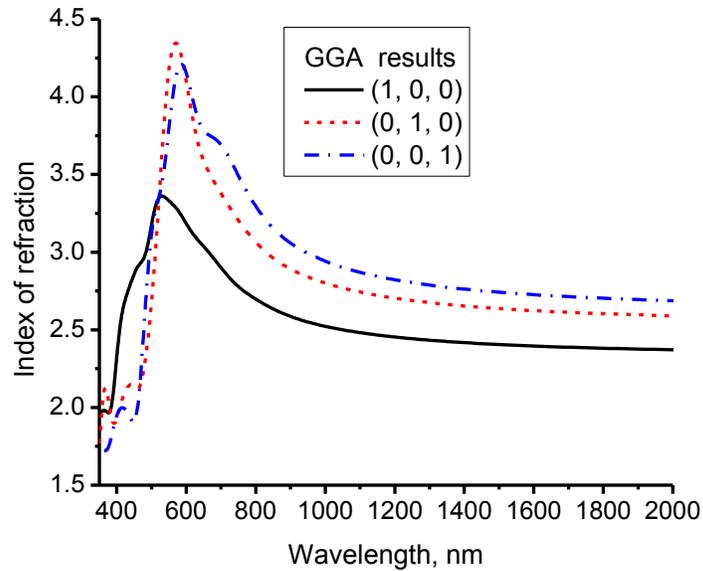

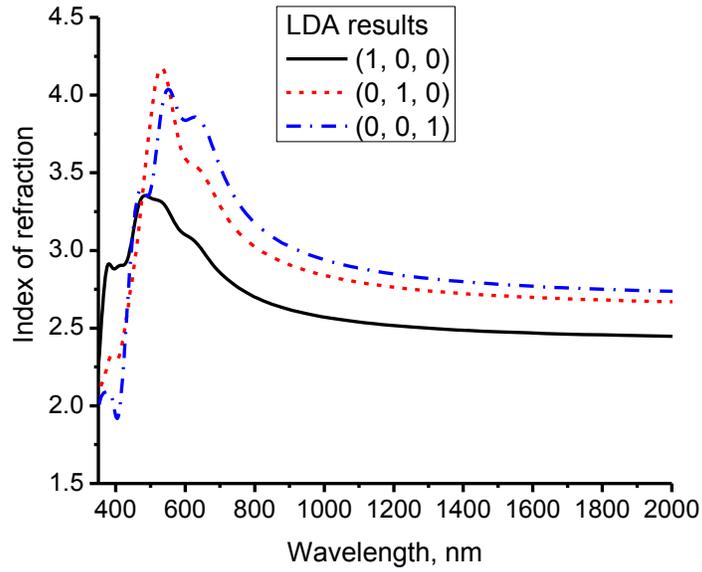

Fig. 7. The GGA- and LDA-calculated refractive index for $CuYS_2$ in different polarizations.

In addition, Fig. 7 presents the calculated dependence of the refractive index on wavelength for three polarizations for $CuYS_2$. Both GGA and LDA results show the same trend: index of refraction has a maximum value (in the long wave length limit) for the (0, 0, 1) polarization, a minimum value for the (1, 0, 0) polarization, and an intermediate value for the (0, 1, 0) polarization. There are the regions of the anomalous dispersion, when the index of refraction increases with increasing wavelength; they are located in the visible part of the spectrum and are due to the interband absorption.

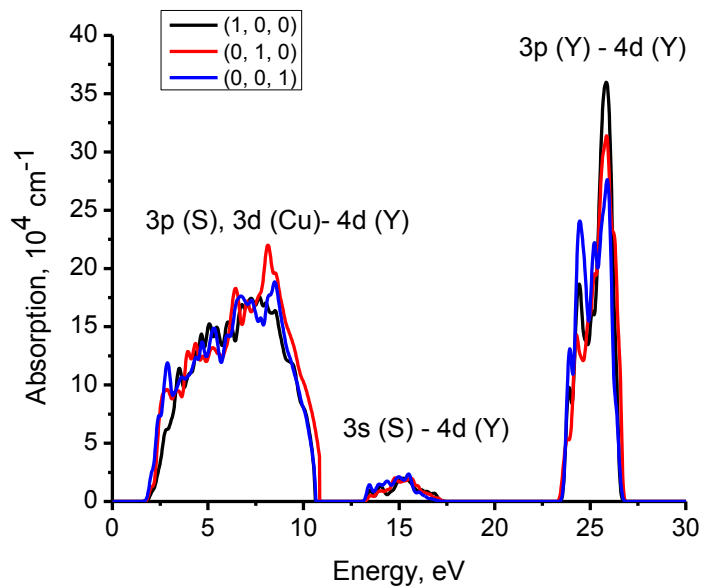

Fig. 8. Calculated absorption spectrum for three different polarizations (indicated in the Figure) for $CuYS_2$.

Fig. 8 shows the calculated absorption spectrum of $CuYS_2$ for three different polarizations in a wider spectral region up to 30 eV. The first and very wide absorption band located between 2 and 11 eV is caused by the transitions between the S 3p, Cu 3d states from the valence band and the Y 4d states in the conduction band. The second absorption band of a much lower intensity, which is centered at about 15 eV, is produced by the transitions between the S 3s states and Y 4d states. Finally, another intensive band at considerably higher energy between 23 and 27 eV arises from the interconfigurational 3p – 4d transitions of yttrium ions. Certain differences in intensities of the calculated absorption bands for different polarizations can be easily noticed in Fig. 8, which emphasizes anisotropic optical properties of $CuYS_2$.

3.4. Elastic properties and pressure effects

Complete set of the calculated elastic constants for $CuYS_2$ is collected in Table 2. No reports on the elastic properties of this compound have been found so far, which does not allow for making comparison with either experimental or theoretical data from other sources.

Table 2. Elastic constants (all in GPa, except for the non-dimensional Poisson ratios $\varepsilon_{ij}$ ($i, j = x, y, z$) and elastic compliance constants $S_{ij}$ (in parenthesis), which are in GPa$^{-1}$)

|  | GGA | LDA |
| --- | --- | --- |
| $C_{11}$ ($S_{11}$) | 102.43 (0.0142169) | 112.46 (0.0126779) |
| $C_{22}$ ($S_{22}$) | 114.64 (0.0108930) | 146.20 (0.0082621) |
| $C_{33}$ ($S_{33}$) | 125.44 (0.0115667) | 143.61 (0.0099983) |
| $C_{44}$ ($S_{44}$) | 25.36 (0.0394272) | 31.14 (0.0321141) |
| $C_{55}$ ($S_{55}$) | 40.59 (0.0246364) | 45.55 (0.0219535) |
| $C_{66}$ ($S_{66}$) | 30.50 (0.0327836) | 28.94 (0.0345521) |
| $C_{12}$ ($S_{12}$) | 42.38 (-0.0029965) | 45.76 (-0.0021620) |
| $C_{13}$ ($S_{13}$) | 59.01 (-0.0055785) | 65.53 (-0.0049882) |
| $C_{23}$ ($S_{23}$) | 46.43 (-0.0026224) | 52.93 (-0.0020585) |
| Bulk modulus $B$ | 70.02 | 79.87 |
| Young's modulus $E_x$ | 70.33 | 78.88 |
| Young's modulus $E_y$ | 91.80 | 121.04 |
| Young's modulus $E_z$ | 86.46 | 100.02 |
| $\varepsilon_{xy}$ | 0.2108 | 0.1705 |
| $\varepsilon_{yx}$ | 0.2751 | 0.2617 |
| $\varepsilon_{xz}$ | 0.3924 | 0.3935 |
| $\varepsilon_{zx}$ | 0.4823 | 0.4989 |
| $\varepsilon_{yz}$ | 0.2407 | 0.2492 |
| $\varepsilon_{zy}$ | 0.2267 | 0.2059 |

Since the studied crystal is orthorhombic, its elastic properties are anisotropic, as can be evidenced by different values of the Young's moduli $E_x$, $E_y$, $E_z$ along crystallographic axes. Additional useful visualization of the elastic anisotropy of a solid comes from a three-dimensional representation of a directional dependence of the Young modulus. In the case of an orthorhombic crystal, such a three-dimensional surface is given by the following expression [15]:

$$\frac{1}{E} = l_1^4 S_{11} + 2 l_1^2 l_2^2 S_{12} + 2 l_1^2 l_3^2 S_{13} + l_2^4 S_{22} + 2 l_2^2 l_3^2 S_{23} + l_3^4 S_{33} + l_2^2 l_3^2 S_{44} + l_1^2 l_3^2 S_{55} + l_1^2 l_2^2 S_{66} , \quad (3)$$

where $E$ is the value of the Young's modulus in the direction determined by the direction cosines $l_1, l_2, l_3$, and $S_{ij}$ are the elastic compliance constants, which form the matrix inverse to the matrix of the elastic constants (Table 2). Application of Eq. (3) to the case of CuYS$_2$ results in the following Young's moduli surfaces and their cross-sections, shown in Fig. 9 (GGA results) and Fig. 10 (LDA results).

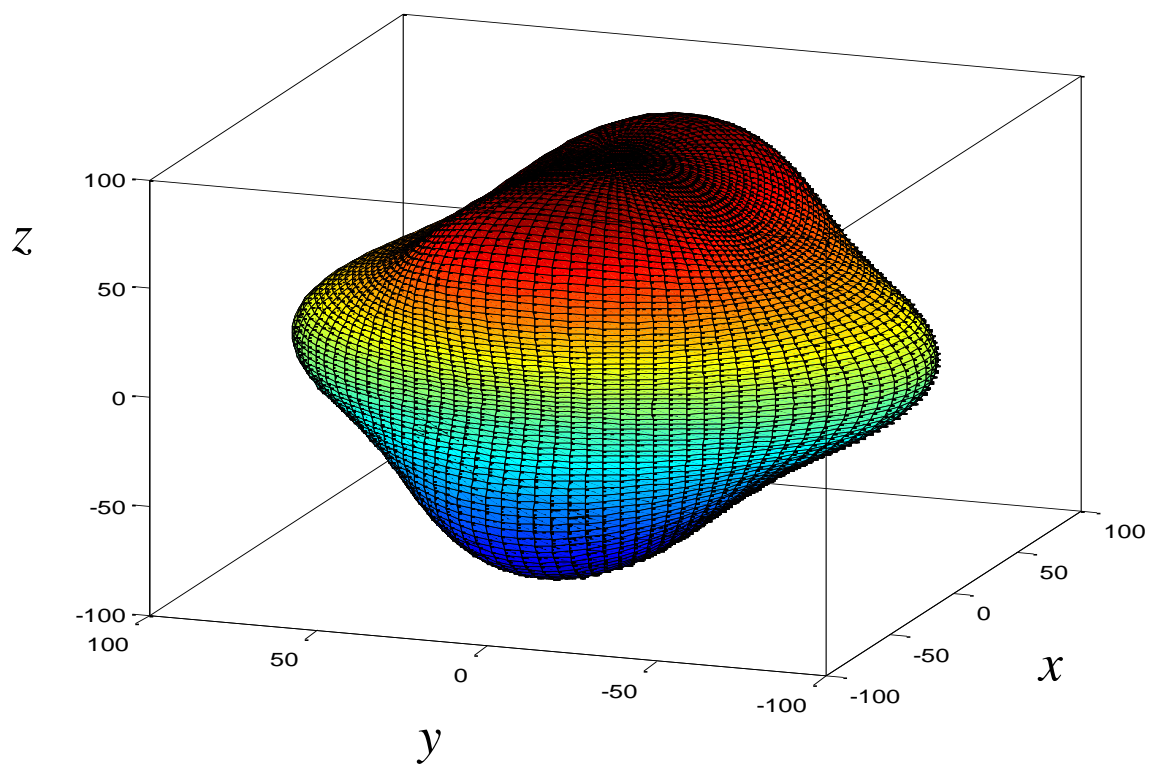

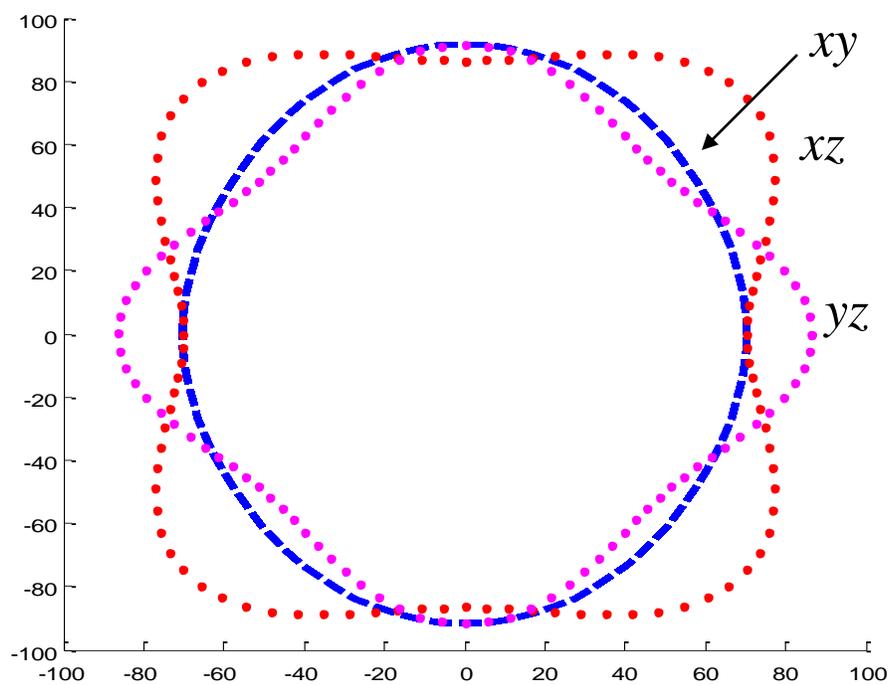

Fig. 9. Young's moduli surface and its cross-sections in the *xy*, *xz*, and *yz* planes for CuYS$_2$ (GGA calculations). The axes units are GPa.

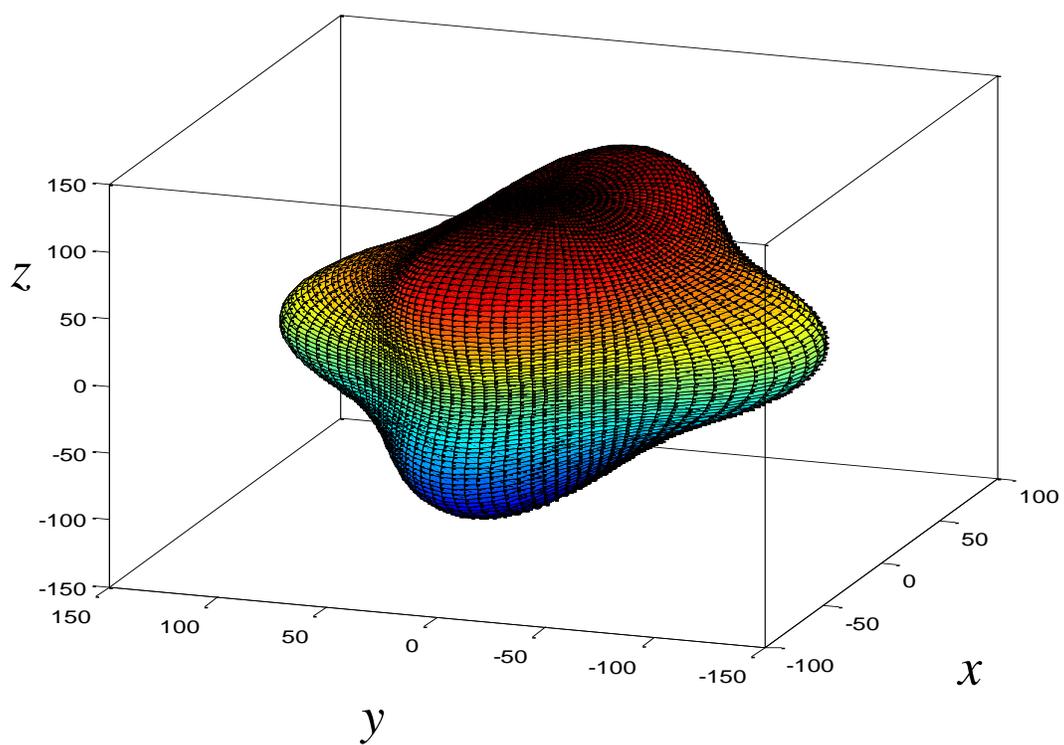

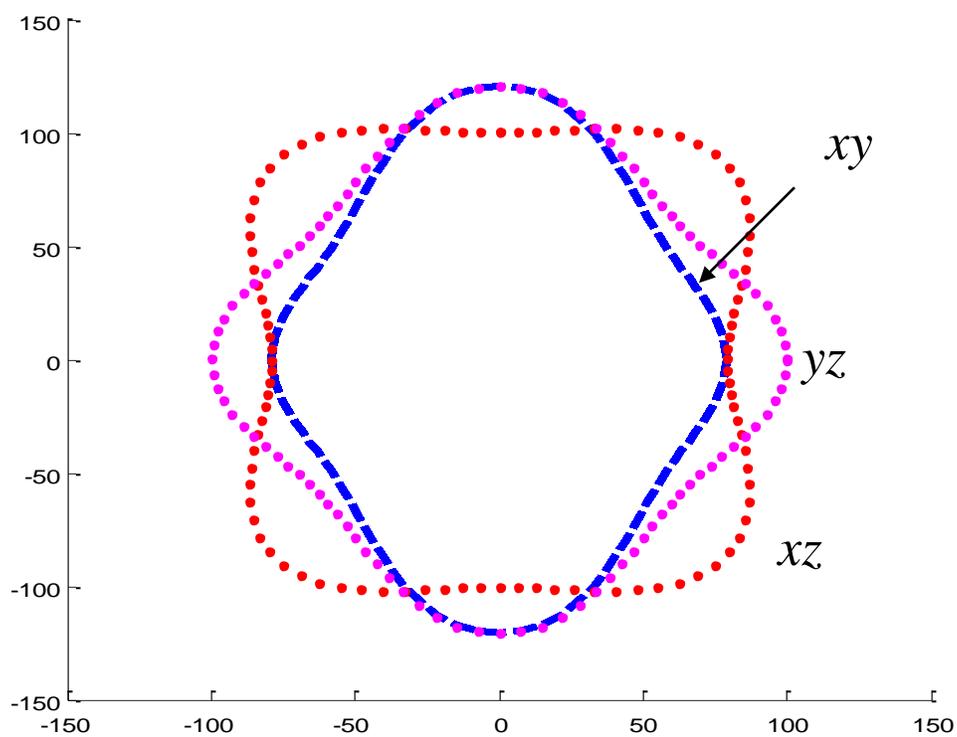

Fig. 10. Young's moduli surface and its cross-sections in the *xy*, *xz*, and *yz* planes for CuYS$_2$ (LDA calculations). The axes units are GPa.

The elastic anisotropy of $CuYS_2$ is clearly visible in Figs. 9 and 10. The cross-sections of the Young's moduli surface look like ellipses in the *xy* plane and like rectangles with rounded corners and depressions at the centers of the sides in the *xz*, *yz* planes. It can be also noticed that for the *yz* cross-section the corners (or the maxima values of the Young's moduli) of those rectangles are on the *y*, *z* axes with the depressions (lowest Young's moduli) located on the diagonals. For the *xz* the situations becomes opposite: the lowest Young's modulus values are realized along the coordination axes, whereas the highest - along the diagonals in the corresponding plane. Since the LDA usually underestimates the lattice constants and, as a consequence, volume of a unit cell, the LDA calculated elastic constants are somewhat grater than their GGA counterparts.

Since the $C_{11}$ constant has the smallest value out of three constants $C_{11}$, $C_{22}$, and $C_{33}$ constants, the crystal lattice of $CuYS_2$ should be more easier compressible along the *a* axis. Compressibility along the b and c axes should be approximately equal, since the values of $C_{22}$ and $C_{33}$ constants are close to each other.

A deeper understanding of the elastic properties of a solid can be gained by performing first principles calculations of its structural and electronic properties at elevated pressure. This has been done for $CuYS_2$ in the pressure range from 0 to 20 GPa with a step of 5 GPa; all calculating settings were the same as described in the earlier sections.

Thus, Fig. 11 presents the calculated lattice parameters *a*, *b*, and *c* (shown by symbols) in their variation with pressure.

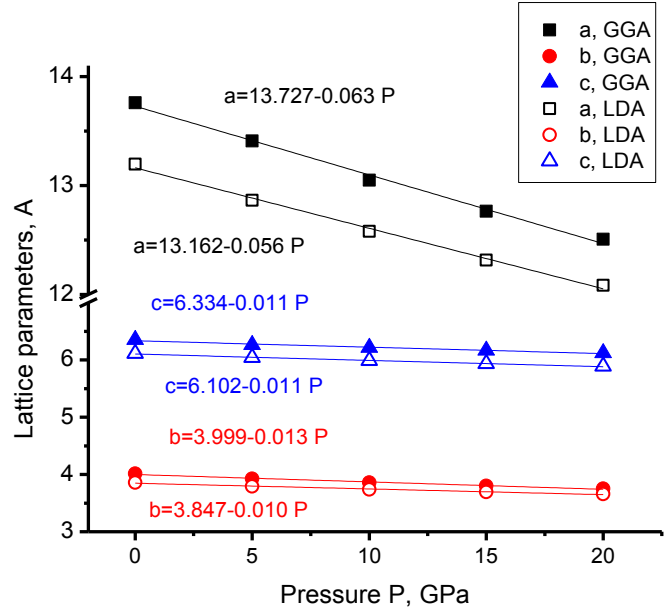

Fig. 11. Pressure dependence of the lattice parameters (filled/open symbols – GGA/LDA) and linear approximations (solid lines). Equations of the linear fits are also given.

As seen from Fig. 11, all three lattice parameters can be pretty well approximated by linear functions, with the slope being equal to the compressibility along a particular crystallographic axis. Those approximating lines for the GGA and LDA calculations are practically parallel, showing consistency of the obtained results. As a general feature of the LDA calculations, they give the absolute values of the lattice parameters, which are slightly smaller than those from the GGA results. The compressibility along the *a* axis is the highest (with the slope of about 0.06 Å/GPa), whereas the compressibilities along the *b* and *c* axes are nearly equal (with the slope of about 0.01 Å/GPa) - this result is in full agreement with the conclusions drawn from the analysis of the elastic constants.

Decrease of the volume of a crystal with pressure can be fitted to the Murnaghan equation of state [16]

$$\frac{V}{V_0} = \left(1 + B'\frac{P}{B}\right)^{-\frac{1}{B'}}, \qquad (4)$$

where $V$ and $V_0$ are the volumes at the elevated pressure $P$ and ambient pressure, respectively, B is the bulk modulus and B' is its pressure derivative. As a rule, the value of B' is between 3 and 4 for solids.

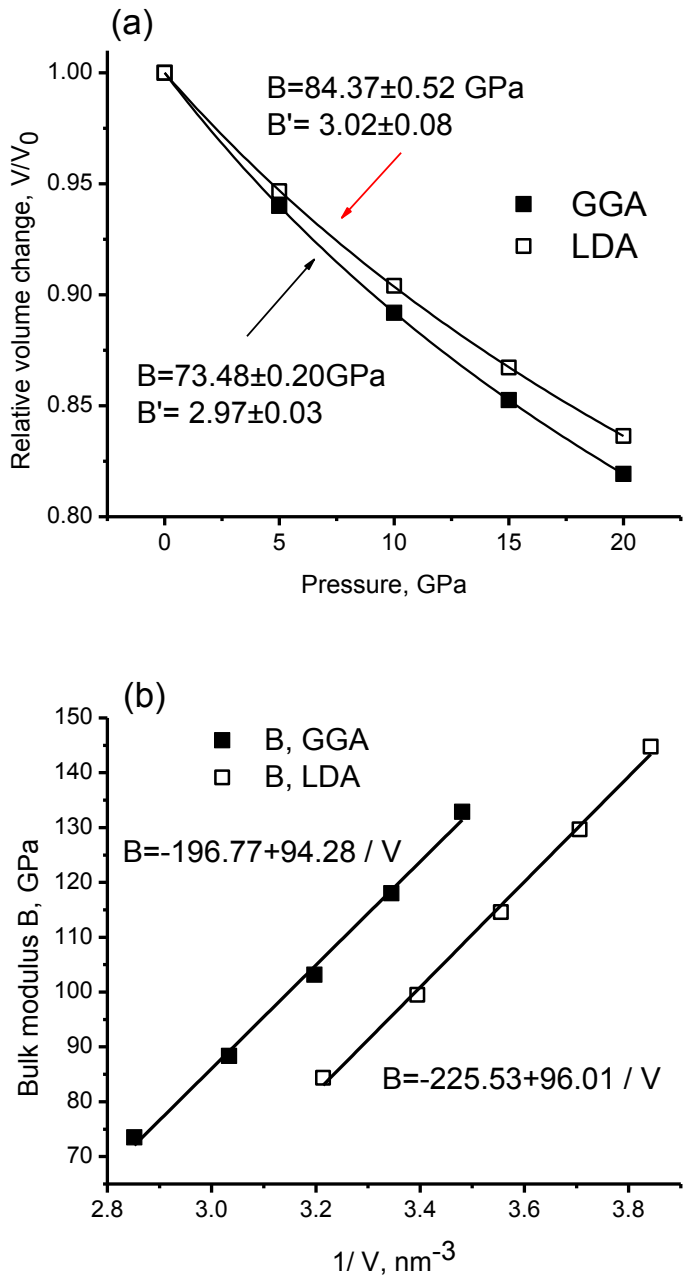

Fig. 12. Calculated dependence of the relative volume $V/V_0$ change (filled/open symbols – GGA/LDA) and the fits to the Murnaghan equation (solid lines) (a) and dependence of the bulk modulus on the inverse volume of the unit cell (b) for $CuYS_2$. The values of the bulk modulus $B$ and its pressure derivative $B'$ are given in Fig. (a).

Fig. 12a shows the relative volume $V/V_0$ change for $CuYS_2$ along with the Murnaghan equation fits. The $B$ values extracted from such a fit (84.37 GPa (LDA) and 73.48 GPa (GGA)) are very close to those obtained from the elastic constants calculations (Table 2). Fig. 12b presents linear dependence of the bulk module on the unit cell volume. The volume of the unit cell was optimized at every value of pressure from the set of 0, 5, 10, 15, 20 GPa; the corresponding values of $B$ were obtained using the $B$ and $B'$ data from Fig. 12a. The bulk modulus $B$ depends on the unit cell volume $V$ as $B = -196.77 + 94.28/V$ (GGA) and $B = -225.53 + 96.01/V$ (LDA); in these two equations $V$ is expressed in nm$^3$ and $B$ in GPa.

Finally, applied hydrostatic pressure also modifies the electronic structure of a solid, leading to an increase (as a rule) of a direct band gap and a decrease (quite often) of an indirect band gap.

Table 3 shows the calculated values of the direct band gaps at the special points of the Brillouin zone (see Fig. 3); the calculated results are also visualized in Fig. 13.

Table 3. Calculated values of the band gaps (in eV) for $CuYS_2$ at varying hydrostatic pressures. The GGA/LDA values are shown.

| Pressure, GPa | G | Z | T | Y | S | X | U | R | Indirect gap |
|---|---|---|---|---|---|---|---|---|---|
| 0 | 1.41/1.55 | 1.885/1.99 | 1.865/1.955 | 1.595/1.78 | 2.77/3.035 | 2.295/2.495 | 2.105/2.275 | 2.305/2.49 | 1.342/1.389 |
| 5 | 1.35/1.54 | 1.83/1.95 | 1.82/1.91 | 1.625/1.81 | 2.895/3.125 | 2.375/2.450 | 2.045/2.36 | 2.28/2.475 | 1.245/1.288 |
| 10 | 1.32/1.52 | 1.785/1.91 | 1.75/1.845 | 1.64/1.875 | 2.975/3.18 | 2.33/2.41 | 1.975/2.06 | 2.25/2.45 | 1.127/1.172 |
| 15 | 1.275/1.51 | 1.705/1.85 | 1.655/1.79 | 1.70/1.975 | 3.075/3.225 | 2.35/2.375 | 1.895/2.025 | 2.25/2.45 | 0.975/1.032 |
| 20 | 1.225/1.47 | 1.625/1.76 | 1.575/1.72 | 1.725/1.99 | 3.075/3.26 | 2.325/2.33 | 1.75/1.87 | 2.225/2.42 | 0.838/0.923 |

As seen from Fig. 13, only at the S and Y points of the Brillouin zone a noticeable increase of the band gap with pressure is observed; for the all remaining points the band gaps depend only slightly on pressure, and the indirect band gap (the lowest band gap out of all calculated) decreases with increasing pressure.

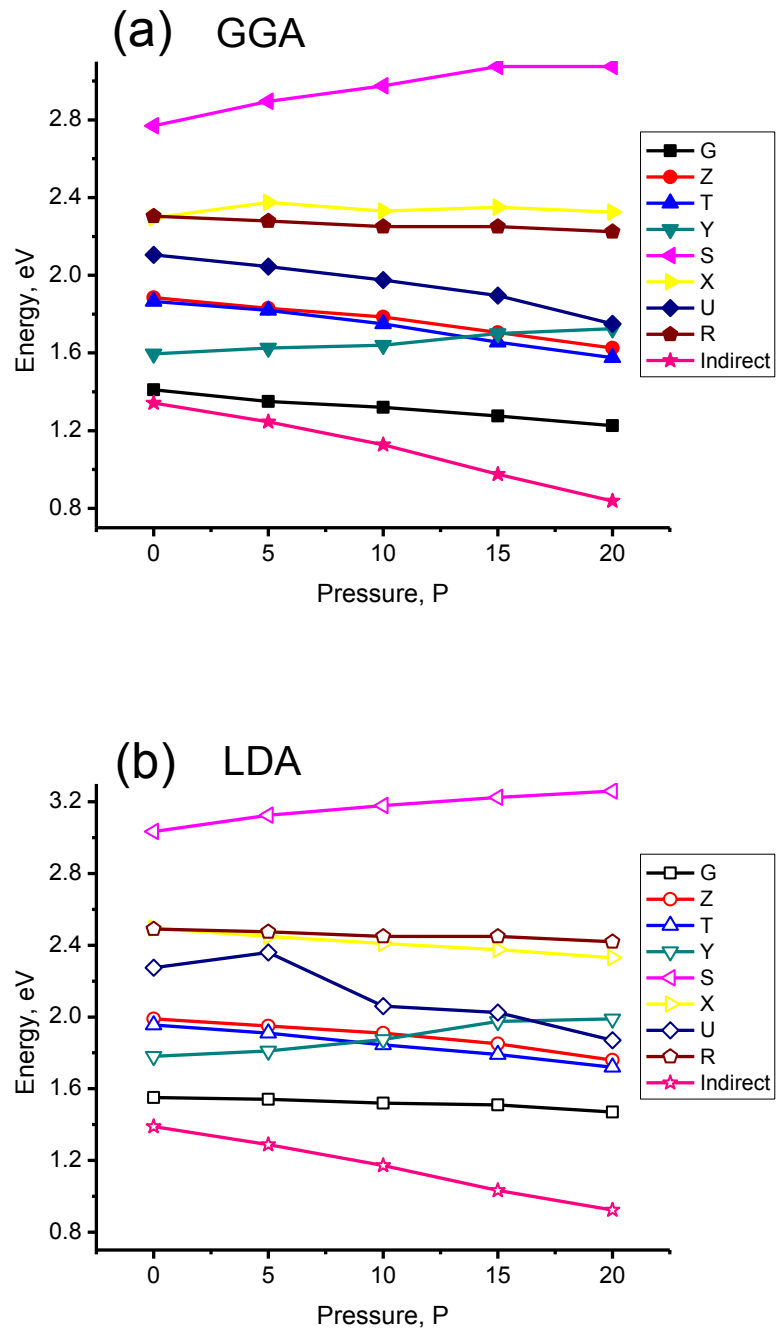

Fig. 13. Calculated band gaps at special points of the Brillouin zone for $CuYS_2$.

## 4. Conclusions

The first principles calculations of physical properties of the ternary semiconductor $CuYS_2$ were performed in the present paper. After successful optimization of the crystal structure, the electronic, optical and elastic properties were all calculated. The indirect band gap was evaluated to be 1.342/1.389 eV (GGA/LDA), the bulk modulus is about 70-80 GPa, depending on the method of calculations.

Since the considered crystal is orthorhombic, its optical and elastic properties should exhibit anisotropy, which was proved by the calculations. In particular, the real and imaginary parts of the dielectric function, absorption spectra and index of refraction were calculated for the (1, 0, 0), (0, 1, 0) and (0, 0, 1) polarizations. Elastic anisotropy was visualized by calculating and plotting the three-dimensional dependence of the Young's moduli on the direction in the $CuYS_2$ crystal lattice and its two-dimensional cross-sections in the *xy*, *yz*, and *xz* planes.

Pressure effects on the lattice parameters and electronic properties were also modeled by optimizing the crystal structure and performing all necessary calculations at elevated hydrostatic pressures. Compressibility along the *a*, *b*, *c* crystallographic axes was calculated and found to correlate with the $C_{11}$, $C_{22}$, and $C_{33}$ components of the elastic tensor.

To the best of the author's knowledge, the results presented in this paper are the first attempt of a first-principles description of the $CuYS_2$ properties.

**Acknowledgment**

The financial support from i) European Social Fund's Doctoral Studies and Internationalisation Programme DoRa and ii) European Union through the European Regional Development Fund (Center of Excellence ''Mesosystems: Theory and Applications'', TK114) is gratefully acknowledged. Dr. A.G. Kumar (University of Texas at San Antonio) is thanked for allowing to use the Materials Studio package.